# On underwater sound reflection from layered ice sheets

Halvor Hobæk[1,2], Hanne Sagen[1],

[1] Nansen Environmental and Remote Sensing Center, Bergen, Norway
[2] Department of Physics and Technology, University of Bergen, Norway
Contact email: halvor.hobak@ift.uib.no

## Abstract

Reflection of sound from ice sheets floating on water is simulated using Thomson and Haskell's [1]-[2] method of matrix propagation. The reflection coefficient is computed as a function of incidence angle and frequency for selected ice parameters of a uniform sheet and two layered ice sheets. At some incidence angles and frequencies the reflection coefficient has very low values. It si shown that this is related to generation of Lamb waves in the ice. The matrix propagation method also provides a dispersion equation for a plate loaded with fluid on one side and vacuum on the other. Finally the concept of beam displacement is briefly discussed.

## 1 Introduction

Propagation of sound waves in Arctic waters is of considerable interest, especially related to long range communication and acoustic thermometry. Due to fresh water from the melting ice and the temperature profile in the upper water layers the sound speed profile tends to form a surface duct, which leads to strong interaction of sound waves with the ice canopy. Thus, the presence of ice and the ice structure bears a strong influence on the under water sound propagation. The structure of the ice canopy is subject to large variations, depending on the history since it was formed. This report concentrates on the simplest of these ice types, namely ice sheets of uniform thickness, possibly of layered structure. The purpose is to provide tables of reflection coefficients to be included in ray tracing models like "RAY" or "Bellhop". If absorption is not taken into account in the ice the reflection is almost total for all frequencies and incidence angles, due to the air interface at the top. However, this is dramatically changed when absorption is included, and strongly influences the reflection coefficient. The main reason for this is excitation of elastic waves, "Lamb waves," in the ice sheet, which due to the absorption is not re-radiated into the water column, as is the case when the waves are not damped.

 This report is mainly based on results from a computation model based on matrix propagation, briefly outlined in the next section. In contrast to analytical models, this model allows the ice sheet to consist of layered ice as for example first year ice and multi year ice (although this is not specifically included in the examples currently presented). In addition, one analytical model, based on [3], is used for comparison, and presented in a separate section. In order to interpret the resulting reflection coefficient map as a function of incidence angle and frequency, it is helpful to introduce the concept of (leaky) Lamb waves. Finally, the reflection coefficient also predicts "beam displacement" which should be accounted for in for example ray tracing models.

## 2 Theory

### 2.1 Matrix propagation

Scattering from a stack of elastic layers is computed by the "matrix propagation method", suggested by Thomson [1] and corrected by Haskell [2]. From a physical point of view the procedure is straightforward: At each interface there is a continuity of displacement and stress, meaning that they are the same at at each side of the interface. Also, if the material properties are constant within each layer, so are the amplitudes of the sound and shear waves except for absorption, which is easily included. Thus, by specifying the displacement-stress components at one interface and relating them to the wave field which subsequently is propagated as plane waves to the next interface, where they are again converted to stress components, the propagation through the stack may be calculated.

The details of this model is presented in [4]. In brief, the waves in the water column or in an ice sheet is specified by a vector of amplitudes of the plane waves representing longitudinal and shear waves. For example in layer $n$ the vector is:

$$\mathbf{\Phi_n} = [\phi_n^-, \psi_n^-, \phi_n^+, \psi_n^+]^T. \tag{1}$$

Here the longitudinal waves are represented by

$$\phi_n(z) = \phi_n^+ \exp(i\alpha_n(z - z_n)) + \phi_n^- \exp(-i\alpha_n(z - z_n)), \tag{2}$$

where we have omitted the factor $\exp(i\xi x - i\omega t)$, which is common to all waves. Here $\xi = \omega/v$, where $v = c/\sin\theta$ is the phase speed of the plane wave at incidence angle $\theta$ in the fluid with sound speed $c$ and angular frequency $\omega$. Thus, $\xi$ is the horizontal wavenumber in the layers. $\alpha_n$ is the vertical wavenumber in the layer ($\alpha = [\omega/c_L]\cos\theta_L$), with longitudinal wave-speed $c_L$ and incidence angle $\theta_L$. "+" indicates waves propagating in the increasing $z$-direction (upwards), and "−" opposite (downwards). The shear waves are similarly expressed as

$$\psi_n(z) = \psi_n^+ \exp[i\beta_n(z - z_n)] + \psi_n^- \exp[-i\beta_n(z - z_n)], \tag{3}$$

where $\beta_n$ is the vertical shear wave number ($\beta = [\omega/c_S]\cos\theta_S$) and $c_S$ is the shear wave speed. The displacement and stress is also specified as a vector

$$\mathbf{S_n} = [u_x, u_z, \sigma_{zz}, \sigma_{zx}]^T, \tag{4}$$

where $u_x$ is horizontal displacement, $u_z$ is vertical displacement, $\sigma_{zz}$ is normal stress and $\sigma_{zx}$ is tangential stress. Note that the order of positioning the potentials and the displacement-stress components vary widely among the literature. The choice used here corresponds with Fuchs [5].

The relation between $\Phi$ and $S$ is found from the basic equations and is represented by a matrix, $\tilde{T}_n$, as given below:

$$\mathbf{S_n} = \tilde{T}_n \cdot \mathbf{\Phi_n}. \tag{5}$$

Denoting the layer of the incident waves as $n = 0$ the displacement/stress vector at the interface to the next layer, at $z = 0$, is found as

$$\mathbf{S_0} = \tilde{T}_0 \cdot \mathbf{\Phi_0} = \mathbf{S_1}, \tag{6}$$

where $\mathbf{S_1}$ is the displacement/stress vector at the other side of the interface. Since $\mathbf{S_1} = \tilde{T}_1 \cdot \mathbf{\Phi_1}$ we find

$$\mathbf{\Phi_1}(z = 0) = \tilde{T}_1^{-1} \cdot \mathbf{S_1} = \tilde{T}_1^{-1} \cdot \tilde{T}_0 \cdot \mathbf{S_0}. \tag{7}$$

The potentials are propagated through the layer by multiplying with $\tilde{E}_1$ to $z = h$:

$$\mathbf{\Phi_1}(z = h) = \tilde{E}_1 \cdot \mathbf{\Phi_1}(z = 0) = \tilde{E}_1 \cdot \tilde{T}_1^{-1} \cdot \tilde{T}_0 \cdot \mathbf{S_0}. \tag{8}$$

Thus, the stress vector at the next interface ($z = h$) becomes

$$\mathbf{S_1}(z = h) = \tilde{T}_1 \cdot \tilde{E}_1 \cdot \tilde{T}_1^{-1} \cdot \tilde{T}_0 \cdot \mathbf{S_0} = \tilde{G}_1 \cdot \tilde{T}_0 \cdot \mathbf{\Phi_0}, \tag{9}$$

where $\tilde{G}_n \equiv \tilde{T}_n \cdot \tilde{E}_n \cdot \tilde{T}_n^{-1}$. The potentials on the other side of this interface then becomes It is now easy to extend this procedure to any numbers of layers, say $N$, such that the semi space after layer $N$ is indexed as $N + 1$.

$$\mathbf{\Phi_{N+1}}(z_{N+1}) = \tilde{T}_{N+1}^{-1} \cdot \tilde{G}_N \cdot \tilde{G}_{N-1} \cdots \tilde{G}_1 \cdot \tilde{T}_0 \cdot \mathbf{\Phi_0} = \tilde{M}\mathbf{\Phi_0}, \tag{10}$$

where

$$\tilde{M} \equiv \tilde{T}_{N+1}^{-1} \cdot \tilde{G}_N \cdot \tilde{G}_{N-1} \cdots \tilde{G}_1 \cdot \tilde{T}_0 \tag{11}$$

Note that since $z$ enters only in $\tilde{E}_n$ (see below) as the distance from the last interface, we may regard this as if the origin of z is reset at each interface.

$$\mathbf{\Phi_2}(z = h) = \tilde{T}_2^{-1} \cdot \tilde{G}_1 \cdot \tilde{T}_0 \cdot \mathbf{\Phi_0}. \tag{12}$$

From the basic wave equations we easily find, with $a_n = e^{i\alpha_n h_n}$, $b_n = e^{i\beta_n h_n}$, (see [4]):

$$\tilde{E}_n = \begin{pmatrix} a_n^{-1} & 0 & 0 & 0 \\ 0 & b_n^{-1} & 0 & 0 \\ 0 & 0 & a_n & 0 \\ 0 & 0 & 0 & b_n \end{pmatrix}, \tag{13}$$

which describes the wave propagation within one layer, and

$$\tilde{T}_n = \begin{pmatrix} i\xi & i\beta_n & i\xi & -i\beta_n \\ -i\alpha_n & i\xi & i\alpha_n & i\xi \\ l_n\mu_n & 2\mu_n\xi\beta_n & l_n\mu_n & -2\mu_n\xi\beta_n \\ 2\mu_n\xi\alpha_n & -l_n\mu_n & -2\mu_n\xi\alpha_n & -l_n\mu_n \end{pmatrix}, \tag{14}$$

which is constant within each layer and independent of $z$ otherwise. Here $\xi$ is horizontal wavenumber, $l = 2\xi^2 - (\omega/c_S)^2$ and $\mu$ is the shear modulus,

We are interested in finding the reflection coefficient for sound waves if the incident waves are propagating in water. This implies that $\tilde{T}_0$ becomes singular, since shear waves are not present. One way to accomplish this is to put $\mu = 0$, meaning that the shear wave speed is equal to zero, but then $\beta$ becomes infinite. The terms with $l\mu$ are unproblematic since it may be shown that they approach $-\omega^2\rho$ when $\mu \to 0$. Likewise terms with $\mu\beta$ may be shown to vanish. However $\beta_0$ remains a problem. One remedy for this is to leave $c_S$ with a small value. It turns out that values with $c_S < 1$ m/s provides good approximation to a fluid.

Since the ice sheet is floating on top of the water column, with air on the top side, reflection must be treated in a special way. The air is substituted with vacuum. It is not meaningful to specify $\Phi_{N+1} = [0,0,0,0]$ for vacuum. Instead one may specify the displacement/stress vector at the vacuum interface, where there are no restrictions on the displacement components, while the stresses $\sigma_{zz}$ and $\sigma_{zx}$ vanish at this interface. Thus,

$$\mathbf{S_{N+1}} = \tilde{G}_N \cdot \tilde{G}_{N-1} \cdots \tilde{G}_1 \cdot \mathbf{S_0} = \tilde{C} \cdot \tilde{T}_0 \cdot \Phi_0. \tag{15}$$

For $\Phi_0$ we put $\Phi_0 = [R, 0, 1, 0]$, and if we define $\tilde{H} = [\tilde{C} \cdot \tilde{T}_0]^{-1}$ we get

$$[R, 0, 1, 0]^T = \tilde{H} \cdot S_{N+1} = \tilde{H} \cdot S_N = \tilde{H} \cdot [u_x, u_z, 0, 0]^T. \tag{16}$$

This results in 4 equations ($h_{ij}$ is element of $\tilde{H}$):

$$R = h_{11} u_x + h_{12} u_z \tag{17}$$
$$0 = h_{21} u_x + h_{22} u_z \tag{18}$$
$$1 = h_{31} u_x + h_{32} u_z \tag{19}$$
$$0 = h_{41} u_x + h_{42} u_z, \tag{20}$$

assuming an incident wave of unit amplitude and a reflection coefficient $R$. If we define $q = u_x/u_z$ Eqs. 18 and 20 gives $q = -h_{22}/h_{21} = -h_{42}/h_{41}$, while Eq. 17 gives $R = u_z(h_{11} \cdot q + h_{12})$, and Eq. 19 $u_z = 1/(h_{31} \cdot q + h_{32})$. Combining this results in

$$R = \frac{-h_{11} h_{22} + h_{12} h_{21}}{-h_{31} h_{21} - h_{32} h_{22}} = -\frac{H^{\Delta}_{1,1}}{H^{\Delta}_{4,1}}, \tag{21}$$

where the first relation for $q$ is used. Here $H^{\Delta}_{1,1}$ is the Delta matrix $h|^{12}_{12}$, etc. (Delta matrices are discussed in [4]). The last relation for $q$ gives $R = H^{\Delta}_{3,1}/H^{\Delta}_{6,1}$, which in practice turns out to be numerically the same.

In order to implement the calculation of $H$ it is necessary to invert $\tilde{C} \cdot \tilde{T}_0$, which gives

$$H = \tilde{T}_0^{-1} \tilde{T}_1 \tilde{E}_1^{-1} \tilde{T}_1^{-1} \tilde{T}_2 \cdots \tilde{T}_N \tilde{E}_N^{-1} \tilde{T}_N^{-1}. \tag{22}$$

Of interest is also the dispersion equation for this situation, which may be found from the same equations. From Eqs. 18 and 20 we see that we must also require

$$\begin{vmatrix} h_{21} & h_{22} \\ h_{41} & h_{42} \end{vmatrix} = 0 \tag{23}$$

which is equivalent to requiring $H^{\Delta}_{5,1} = 0$. This gives the required dispersion equation for (leaky) Lamb waves in the ice sheet loaded by water on one side and vacuum on the other.

## 2.2 Analytical model for reflection from an homogeneous ice sheet

According to Brekhovskikh [3] the reflection coefficient of a plane wave at the ice-water interface is given by

$$R = \frac{-(MZ_1)Z_3 + i[(MZ_1)^2 - (NZ_1)^2]}{(MZ_1)Z_3 + i[(MZ_1)^2 - (NZ_1)^2]}, \tag{24}$$

where
$$MZ_1 = Z_2 \cos^2 2\gamma \cot p + Z_{2t} \sin^2 2\gamma \cot Q$$
$$NZ_1 = Z_2 \cos^2 2\gamma / \sin p + Z_{2t} \sin^2 2\gamma / \sin Q$$
$$Z_2 = \rho_{ice} c_L / \cos\theta_L, \; Z_{2t} = \rho_{ice} c_S / \cos\gamma,$$
$$Z_3 = \rho c / \cos\theta,$$
$$p = kh\cos\theta_L, \; Q = Kh\cos\gamma,$$
$$k = \omega/c_L, \; K = \omega/c_S,$$
$$\frac{\sin\theta}{c} = \frac{\sin\gamma}{c_S} = \frac{\sin\theta_L}{c_L}.$$

Here $\rho_{ice}$, $c_L$, $c_S$ are the density, compressional wave and shear wave speeds in the ice. The subscripts $L$ and $S$ denote the compressional and shear wave, respectively, in the ice layer. No subscript is used for sea water parameters, $h$ is the ice thickness. $\theta$ is the incident angle. $\theta_L$ and $\gamma$ are the refraction angles for the compressional and shear waves, respectively, in the ice layer. The ice sheet consists of only one layer, bounded by vacuum (air) at the top.

A Matlab program for computing the reflection coefficient using this model is included in the Appendix.

## 2.3 Lamb modes

The dispersion equation for Lamb waves in a plate, with and without fluid loading (on both sides), were presented by Shock [6]. In terms of the abbreviations $s = (c_S/v)^2$, $q = (c_S/c_L)^2$, $r = (c_S/c)^2$ and the ratio of fluid density $\rho$ and layer density $\rho_L$, two sets of modes appear:

$$h_s = (1-2s)^2 \cot(\alpha h/2) + 4s\sqrt{(1-s)}\sqrt{(q-s)} \cot(\beta h/2) - i\frac{\rho\sqrt{(q-s)}}{\rho_L\sqrt{(r-s)}} = 0, \quad (25)$$

and

$$h_a = (1-2s)^2 \tan(\alpha h/2) + 4s\sqrt{(1-s)}\sqrt{(q-s)} \tan(\beta h/2) + i\frac{\rho\sqrt{(q-s)}}{\rho_L\sqrt{(r-s)}} = 0, \quad (26)$$

where $h_s$ represents "symmetrical modes", and $h_a$ "asymmetrical modes".

Free Lamb mode waves in an unloaded plate is found by putting the fluid density $\rho = 0$, and the equations simplify to

$$\left(\frac{\tan\beta h/2}{\tan\alpha h/2}\right)^{\pm 1} = -4\frac{\alpha\beta\xi^2}{(\beta^2 - \xi^2)^2}. \quad (27)$$

The exponent $+1$ represents symmetrical modes, $-1$ asymmetrical.

A Matlab program for computing $h_s$ and $h_a$ is shown in the Appendix. The modes are found by selecting locations of $v$ and frequency giving amplitudes of $h_s$ and $h_a$ less than a given threshold. Unfortunately, the fluid loading on both sides assumes a symmetry which is broken when the loading is only on one side. It is not an option to reduce the fluid loading by a factor 2. Thus, for the case at hand, the equations with fluid loading are not very relevant.

### 2.4 Beam displacement

According to Brekhovskikh [3] there will be a displacement of a "sound beam" along the interface depending on the variation of the phase of the reflection coefficient with the horizontal wave number (i.e. the component along the interface), such that

$$\Delta = -\frac{\partial \phi}{\partial \xi}, \tag{28}$$

where $\phi$ is the phase angle of the reflection coefficient, such that $R = |R| \exp i\phi$. Since $\xi$ depends on both the frequency and the incidence angle: $\xi = \omega \sin\theta / c_f$ where $c_f$ is sound speed and $\theta$ is incidence angle in the fluid, we get:

$$d\xi = \frac{\partial \xi}{\partial \omega} d\omega + \frac{\partial \xi}{\partial \theta} d\theta = \frac{\sin\theta}{c_f} d\omega + \frac{\cos\theta}{c_f} \omega d\theta. \tag{29}$$

Thus,

$$\frac{\partial \phi}{\partial \xi} = \frac{\partial \phi}{\partial \theta}\frac{d\theta}{d\xi} + \frac{\partial \phi}{\partial \omega}\frac{d\omega}{d\xi} = \frac{\partial \phi}{\partial \theta}\frac{c_f}{\omega \cos\theta} + \frac{\partial \phi}{\partial \omega}\frac{c_f}{\sin\theta}, \tag{30}$$

giving

$$\Delta = -c_f \left[\frac{\partial \phi}{\partial \theta}\frac{1}{\omega \cos\theta} + \frac{\partial \phi}{\partial \omega}\frac{1}{\sin\theta}\right], \tag{31}$$

or, in terms of the frequency $f$:

$$\Delta = -\frac{c_f}{2\pi}\left[\frac{\partial \phi}{\partial \theta}\frac{1}{f \cos\theta} + \frac{\partial \phi}{\partial f}\frac{1}{\sin\theta}\right]. \tag{32}$$

In order to use the computed reflection coefficient it is necessary to "unwrap" the phase before computing the derivatives. While applying this expression to the computed reflection coefficient it appears that the last term, representing the derivative with respect to frequency, contributes with very large negative displacements near normal incidence. If the sound beam has a narrow frequency bandwidth this considered to be unrealistic [7]. Examples are shown with only the first term in Eq. 32, and with both terms.

## 3 Ice parameters

Sea ice appears in a wide variety of shapes and structures, such as first year ice (relatively isotropic), multi year ice (much more structured), ice-bergs, and so on, and parameter variation with season (winter/summer). Reference [8] includes a table of ice parameters reported in previous literature. In this table $c_L$ varies between $3000 - 3600$ m/s, $c_S$ between $1500 - 1800$ m/s, and absorption parameters $0.07 < a_L < 0.76$ dB/$\lambda$ and $0.05 < a_S < 2.5$ dB/$\lambda$. The density is typically about 940 kg/m$^3$. In order for the ice to have a positive compressibility it is required that

$$c_S < \sqrt{\frac{3}{4}} c_L, \tag{33}$$

Furthermore, to assure that absorption dissipates energy, not generates it, one finds the requirement

$$\frac{a_S}{a_L} < \frac{3}{4}\left(\frac{c_L}{c_S}\right)^2 \tag{34}$$

Not all the cases cited in [8] satisfy this condition.

In the examples used here 3 different sets of ice parameters are used, as shown in Table 1 (T22 is from Rajan [9]). In all cases the water sound speed is 1500 m/s, and $\rho$=1000 kg m$^{-3}$, and vacuum as the final layer above the ice. The total ice sheet thickness is 5 m for case T4 and T21, and 3 m for case T22. Rajan [9] does not provide values for the shear wave attenuation. It has been chosen to comply with Eq. 34.

| Param. | T4 | T21 | | T22 | | | | | |
|---|---|---|---|---|---|---|---|---|---|
| | | Lower | Upper | Lower | | | | | Upper |
| $c_L$, m/s | 3200 | 3500 | 3850 | 2800 | 3400 | 3600 | 3600 | 3700 | 3900 |
| $c_S$, m/s | 1600 | 1850 | 2040 | 1600 | 1600 | 2000 | 1700 | 1800 | 1900 |
| $\rho$, kg m$^{-3}$ | 930 | 900 | 900 | 905 | 900 | 900 | 890 | 890 | 895 |
| $a_L$, dB/$\lambda$ | 0.1 | 0.1 | 0.1 | 0.194 | 0.235 | 0.249 | 0.249 | 0.256 | 0.270 |
| $a_S$, dB/$\lambda$ | 0.2 | 0.2 | 0.2 | 2*$a_L$ | 2*$a_L$ | 2*$a_L$ | 2*$a_L$ | 2*$a_L$ | 2*$a_L$ |
| h, m | 5 | 3 | 2 | 0.4 | 0.1 | 0.5 | 1.0 | 0.5 | 0.5 |

Table 1: Parameters of ice sheets

### 3.1 Absorption

Absorption is included in the models putting the compressional and shear wave speeds complex:

$$c = c_0(1 - i\delta). \tag{35}$$

Thus, $k = k_0 \frac{1+i\delta}{1+\delta^2}$, which approximates to $k \approx k_0(1 + i\delta) = k_0 + i\alpha$ provided $\delta \ll 1$, with $\alpha = k_0 \delta$. Here $\alpha$ is in units Neper/m. When the absorption is specified in dB/wavelength it is related to $\alpha$ by

$$a^{(\lambda)} = \alpha \lambda 20 \log e = k_0 \delta \lambda 20 \log e = 2\delta\pi 20 \log e,$$

Thus,

$$\delta = \frac{a^{(\lambda)}}{40\pi \log e} = \frac{a^{(\lambda)}}{54.575}. \tag{36}$$

## 4 Results

### 4.1 Reflection coefficient

A typical map of the magnitude of the reflection coefficient for parameters T4 is shown in Figure 1, mapped versus incidence angle and frequency. Note that the frequency axis may be scaled as frequency times thickness. Here $h = 5$ m, so if it were only 1 m, the coefficient along $f = 100$ Hz in the shown plot would correspond to 500 Hz at $h = 1$ m.

Evidently the reflection coefficient depends on both incidence angle and frequency, but it has a magnitude close to 1 in most of the map, except near several dips where it is substantially lower. The dips are labelled in order to compare with the dispersion plot in Figure 5. It is interesting to note that if absorption is excluded the dips become much less dominant, and the reflection coefficient is 1 almost everywhere, as shown in the contour plot in Figure 2. One might note that dip B moves towards 90 degrees incidence angle when $c_S$ is reduced.

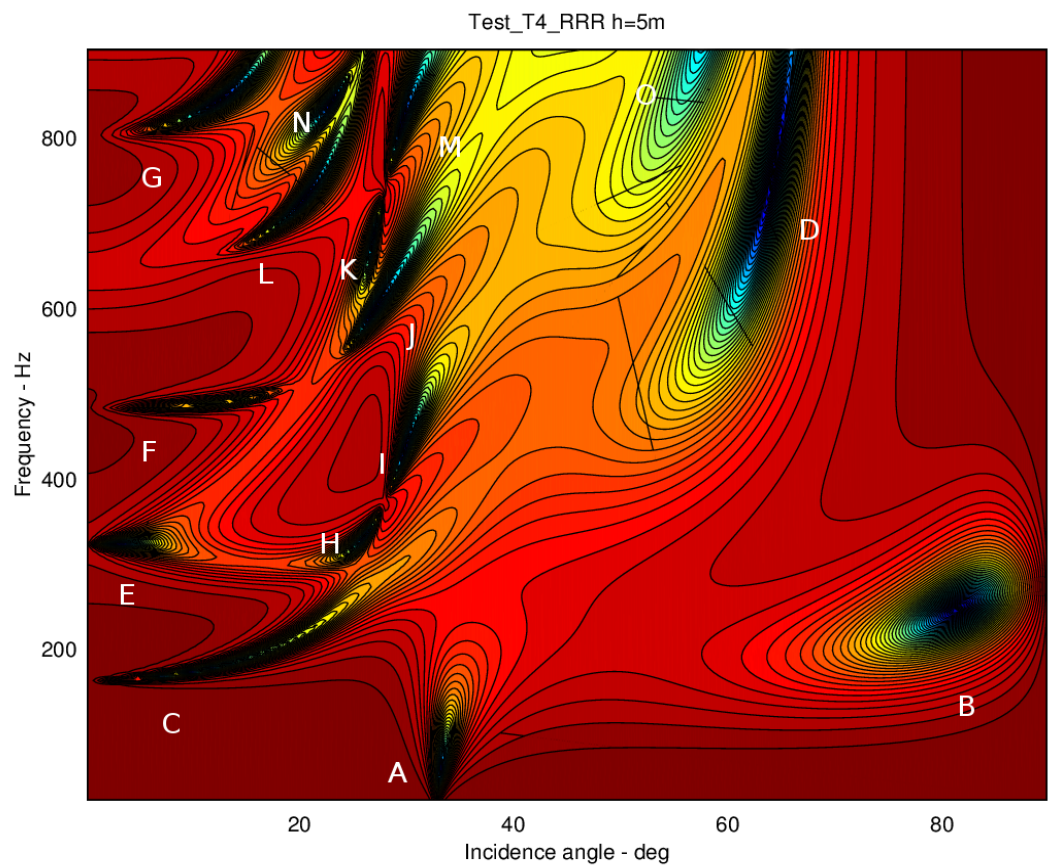

Figure 1: Reflection coefficient for example T4: $c_L$ = 3200 m/s, $c_S$ = 1600 m/s, $\rho$ = 930 kg/m$^3$, $a_L$ = 0.1 dB/$\lambda$ and $a_S$ = 0.2 dB/$\lambda$. Characteristic dips are marked with capital letters. The deep red color indicates magnitude 1.

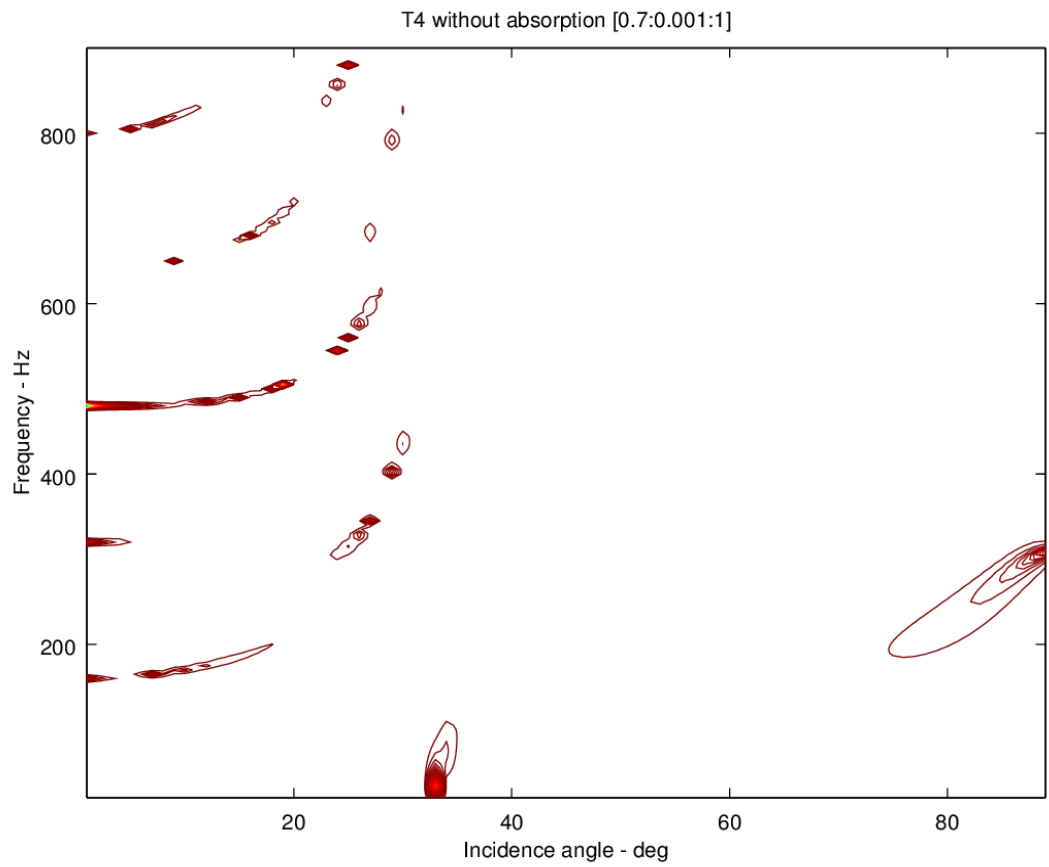

Figure 2: Reflection coefficient for T4 without absorption. Contour plot: white indicates magnitude 1.

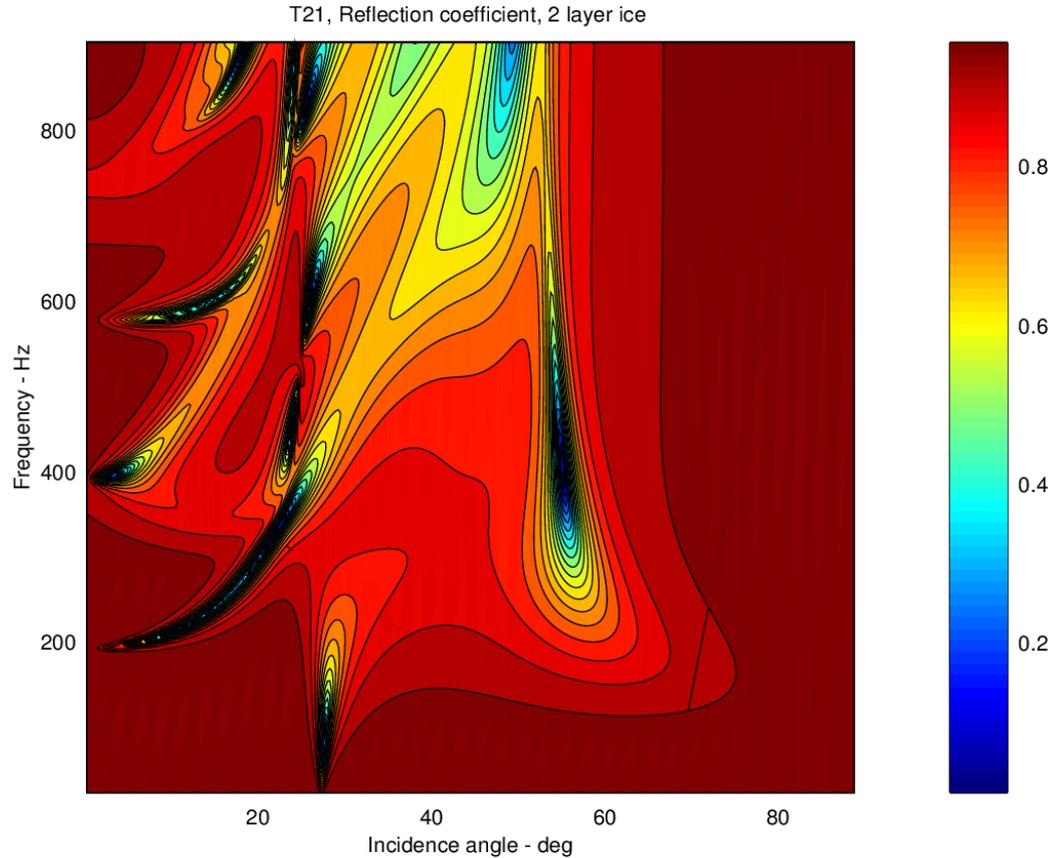

Figure 3: Reflection coefficient for T21, consisting of 2 layers of ice. Note that near grazing incidence the reflection coefficient is close to 1 for all frequencies.

The reflection coefficient of T21 is shown in Figure 3. Note that near gracing incidence the reflection coefficient is much more regular at all frequencies, and close to magnitude 1. Thus, the actual ice consistency is quite important.

Figure 4 shows the reflection coefficient magnitude for T22.

### 4.2 Dispersion diagram

In an attempt to explain the dips the same data is plotted as a map with horizontal phase velocity ($v$) versus frequency in a modified dispersion diagram as shown in Figure 5. The incidence angle is converted to phase speed by $v = c/\sin\theta$. Clearly no phase speeds can be less than $c$ (1500 m/s). The amplitude scale is as in Fig. 11.

Superposed on this plot are also "free" Lamb modes computed from Eq. 27. The symmetric modes are white, the asymmetrical blue, and identified with symbols s and a, respectively. The horizontal blue line at $v = c_L = 3200$ is an artefact. Many of he dips marked in Fig. 11 are identified. Even if the Lamb modes are computed assuming no fluid loading, it is clear that they match the various dips to a large extent. Exceptions are dip B and D, for which we presently have no obvious explanation. It is possible that B is related to mode $s_0$ and D to mode $a_1$ if these modes deform when loaded by fluid on only one side of the plate.

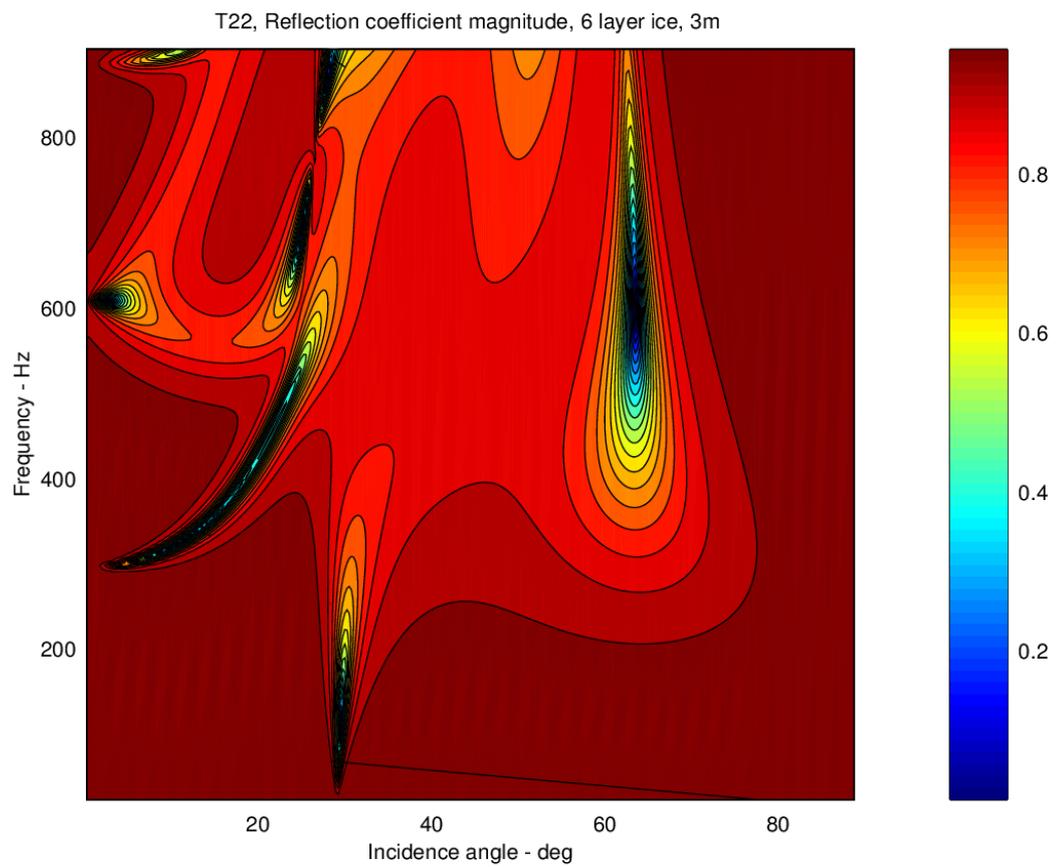

Figure 4: Reflection coefficient for T22, consisting of 6 layers of ice. Since the ice thickness is only 3m the frequency axis is scaled compared to the previous maps. Also here the near grazing angles have reflection coefficients close to total.

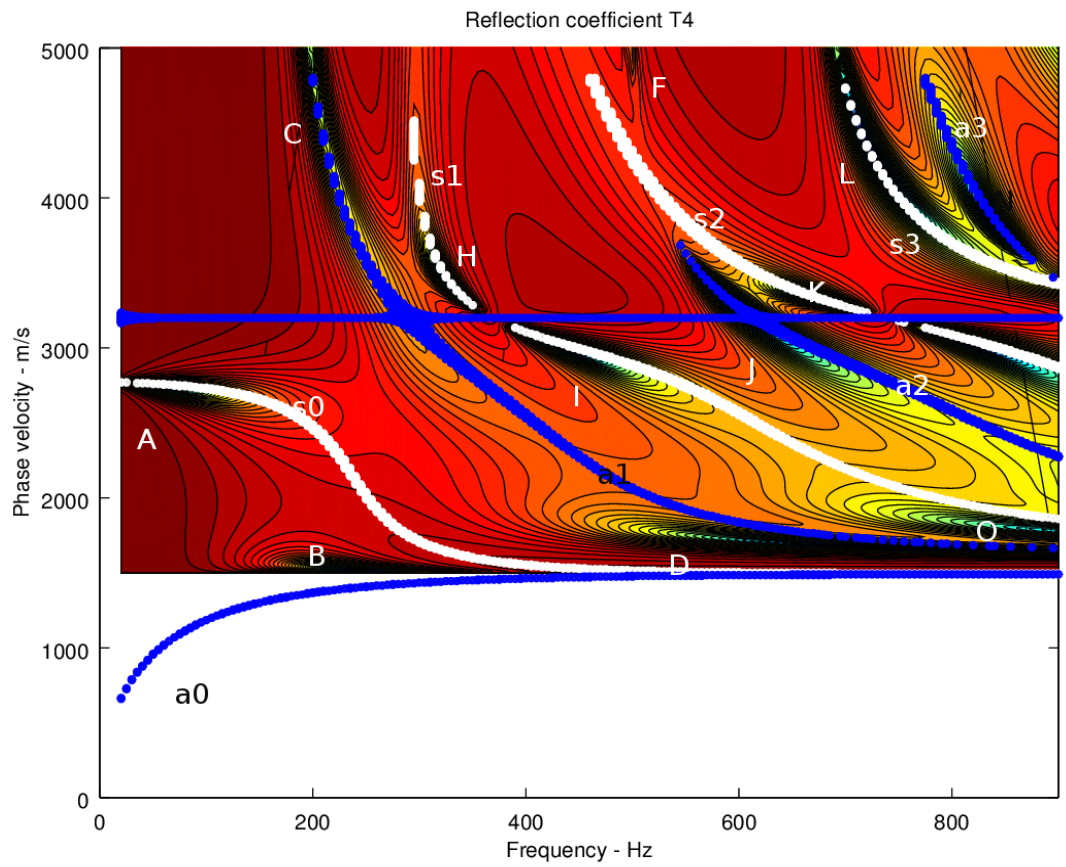

Figure 5: Phase speed of Lamb modes as a function of frequency for the same case as in Figure 1. The dips are the same as marked in Figure 1.

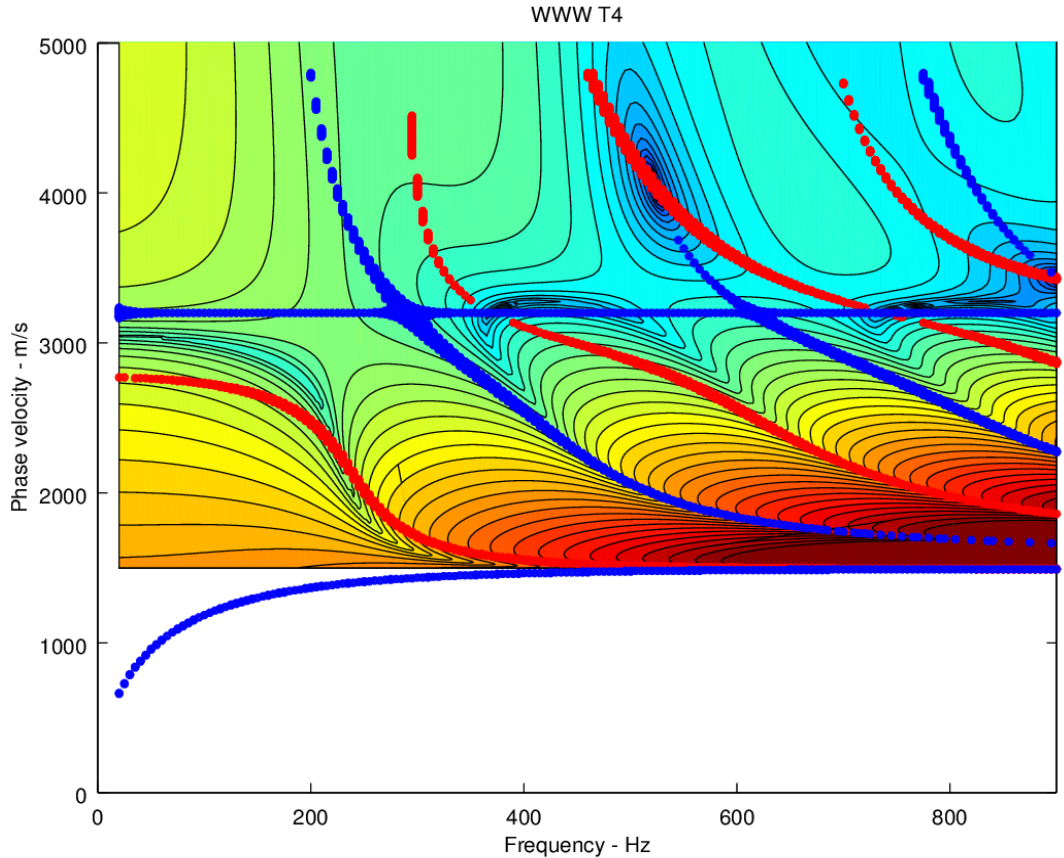

Figure 6: Location of "Lamb modes" based on Eq. 23, superposed "free" Lamb modes computed from Eq. 27.

The main conclusion from these results are that the variations in reflection coefficient is due to generation of Lamb waves in the ice sheet at certain frequencies and incidence angles.

Computations of the "Lamb modes" using Eq. 23 do not match those computed from Eq. 27. Figure 6 show the contour plot of the determinant in Eq. 23. The Lamb modes are located at the minima seen in this map. None of the modes can be identified at phase speeds above $c_L$ m/s.

It is striking that mode "s0" terminates at a phase speed about 3000 m/s at low frequency while the "free" "s0" mode terminates near 2800 m/s, and follows a more gentle slope with increasing frequency. The other modes seem to coincide with free Lamb modes at increasing frequency, and there seems like the "Lamb modes" are connected near phase speeds of $c_L$. Thus, $A_1$ seems to join with $S_1$ and $A_2$ with $S_2$ (capital letters indicate "Lamb modes" in contrast to "free Lamb modes"). The significance of the dips in this plot is not understood. They are not related to dips seen in the reflection coefficient map. On the other hand, they seem to occur where a symmetric and an asymmetric free Lamb mode cross.

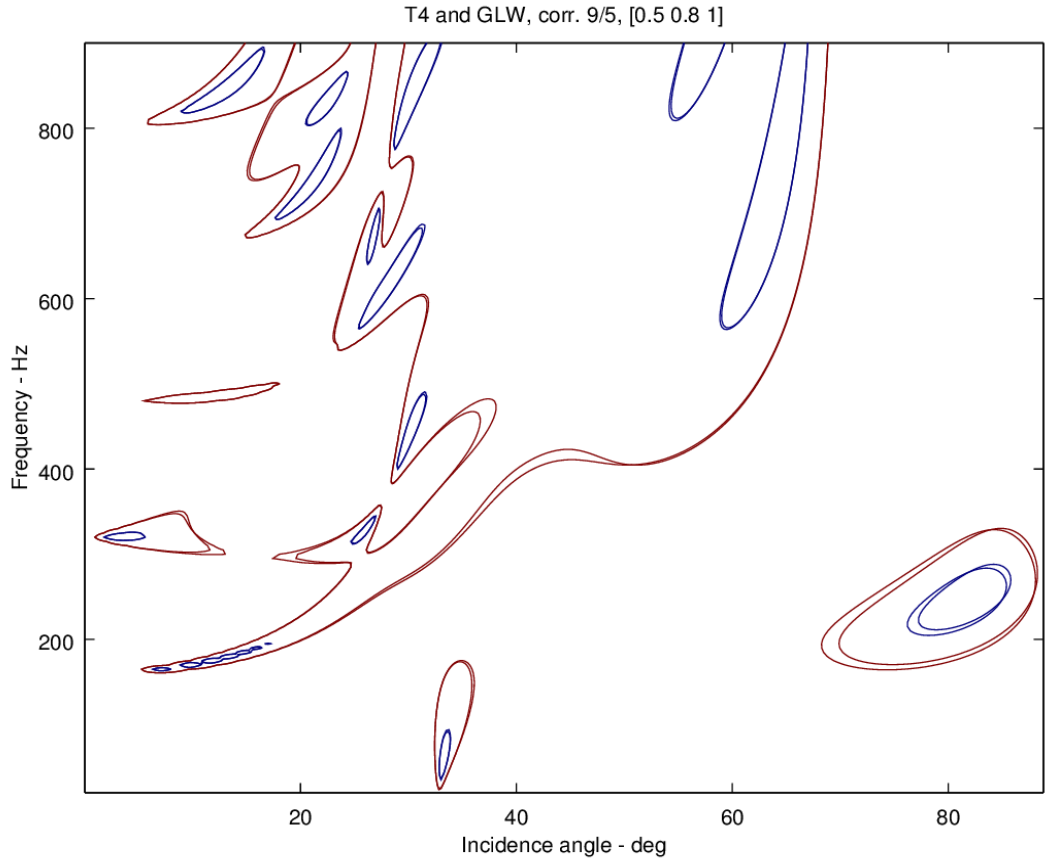

Figure 7: Comparison of T4 contour plots of matrix propagation method and Eq. 24. Contours at $0.5$ and $0.8$.

### 4.3 Comparison to Brekhovskikh's solution

When absorption is included the Brekhovskikh solution, Eq. 24, gives almost identical results to the matrix propagation model, when the ice is not layered. This is exemplified in Figure 7, showing a map of the T4 case computed by both models, and presented as a superposition of contour maps with contours at 0.5 (brown) and 0.8 (blue). Where only one contour is visible they both overlap. However, when absorption is excluded the Brekhovkskikh equation results in reflection coefficient = 1 all over, not even the small dips seen in Figure 2.

### 4.4 Beam displacement

The phase map of T4 is shown unwrapped in Figure 8. Obviously it is rather complicated, and not straightforward to unwrap. In practice this is done along the incidence angle and frequency separately, to obtain the two terms in Eq. 32. As mentioned earlier the frequency term is considered unrealistic in an approximately monofrequency case, since it generates very large negative displacements at near to normal incidence (see also [7]). Figure 9 shows the resulting beam displacement (in m) for T4 using only the first term in Eq. 32. For most of the incidence angles the beam displacement is small (of the order of a few meters), but in some regions it is quite large, typically near the dips seen in the reflection coefficient map. In particular, below abut 200 Hz it exceeds 100 m at grazing incidences. The displacement due to the frequency term is small in these regions, and unable to reduce the displacements significantly, as demonstrated in Figure 10.

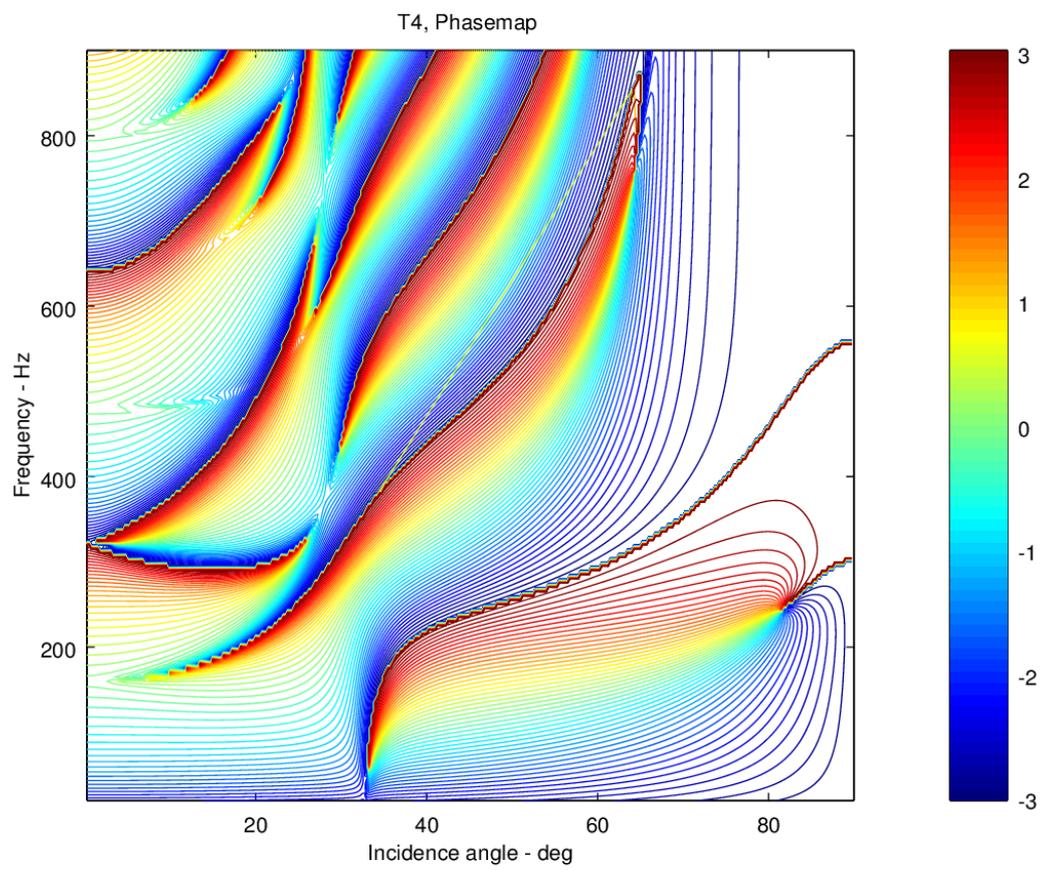

Figure 8: Map of the phase of the reflection coefficient for T4. The contours are in radians.

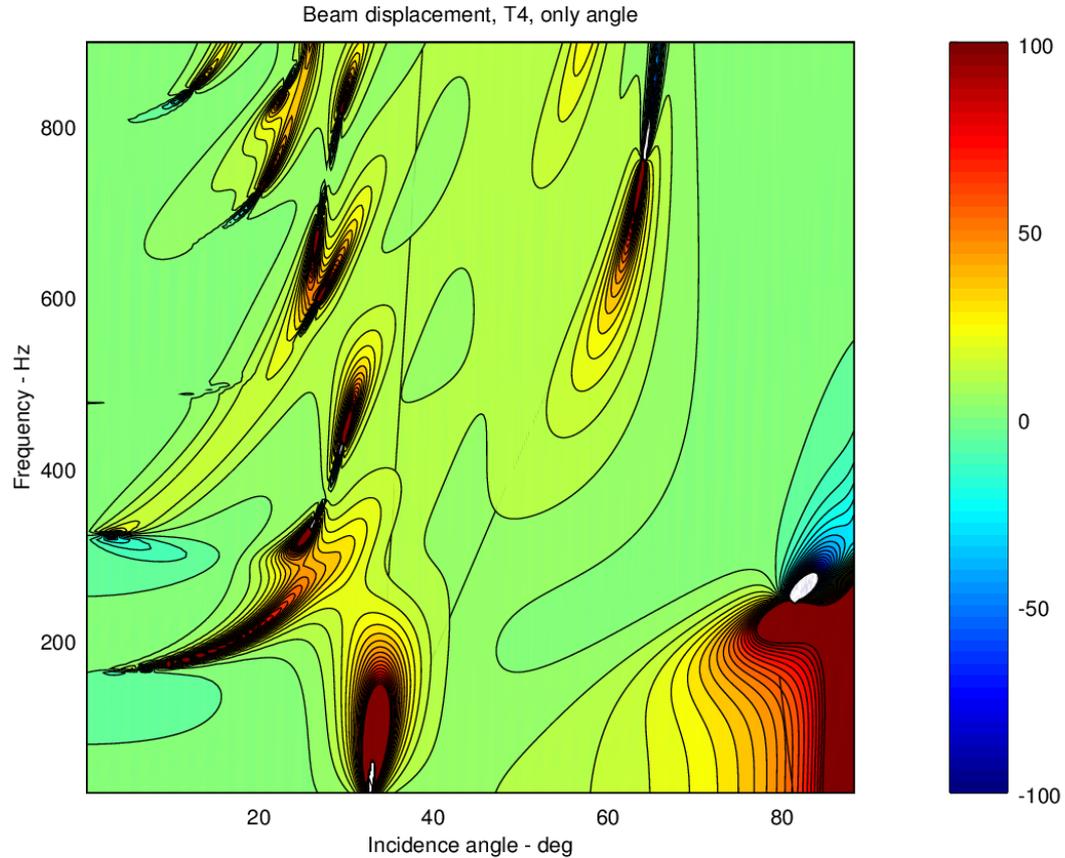

Figure 9: T4 beam displacement (m), only angle term.

Finally, Figure 11 shows the displacement when both terms are included. Apart from a few regions near the dips the displacement is negative. The white regions represent displacements larger than -500 m. It is difficult to find physical reasons for such displacements.

## 5  Discussion and Summary

The matrix propagation method was previously [4] demonstrated as a useful tool for computing the reflection coefficient for a layered bottom, and was here modified to operate on a sheet of ice floating on the water column with air (vacuum) on top. The ice may consist of any number of layers of different material properties. In contrast to the alternative method of "Global Matrix Method" the matrix propagation method cannot tolerate a fluid layer between solid layers, but this case may be approximated by assuming a very low but finite shear wave speed in the fluid layer. This was also used in the water column in this case. For the case of a homogeneous ice sheet (i.e. only one layer) the comparison with an analytical model by Brekhovskikh [3] yields almost identical results. Since it previously was shown to give good results for a layered bottom, the matrix propagation model should be approved for computing reflection coefficients from layered ice.

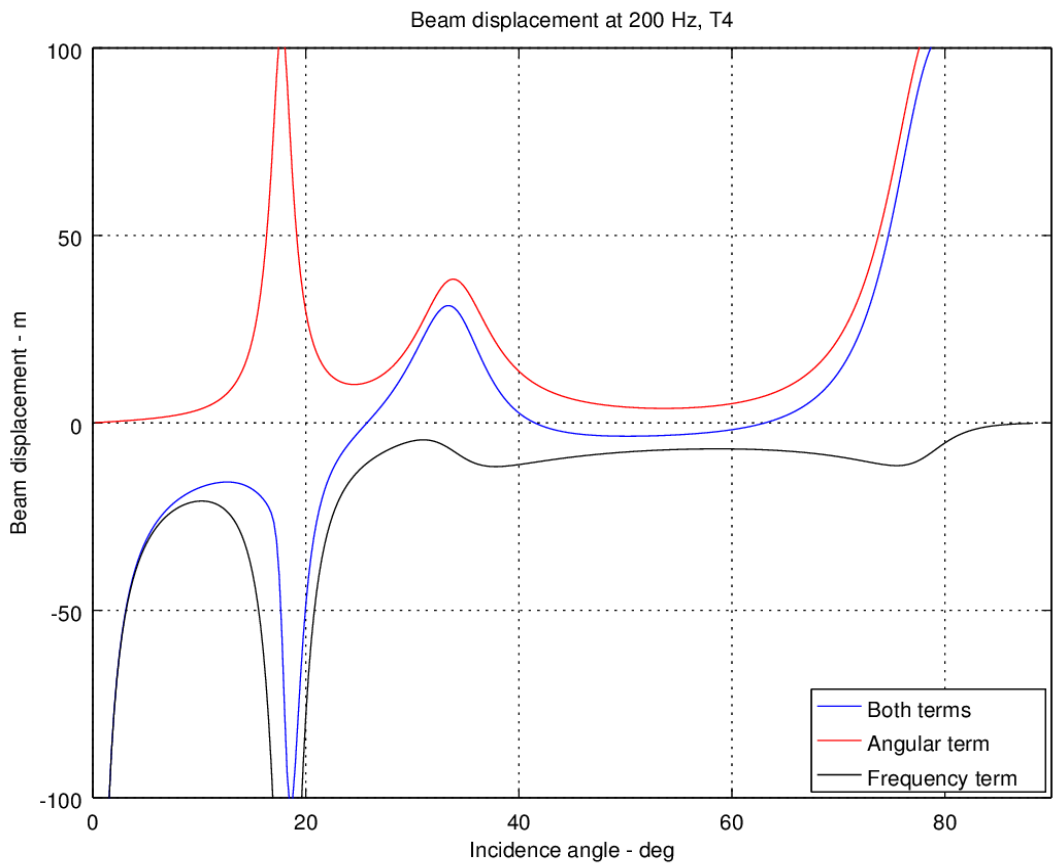

Figure 10: T4, Beam displacement at 200 Hz. The two terms in Eq. 32 are shown individually and summed.

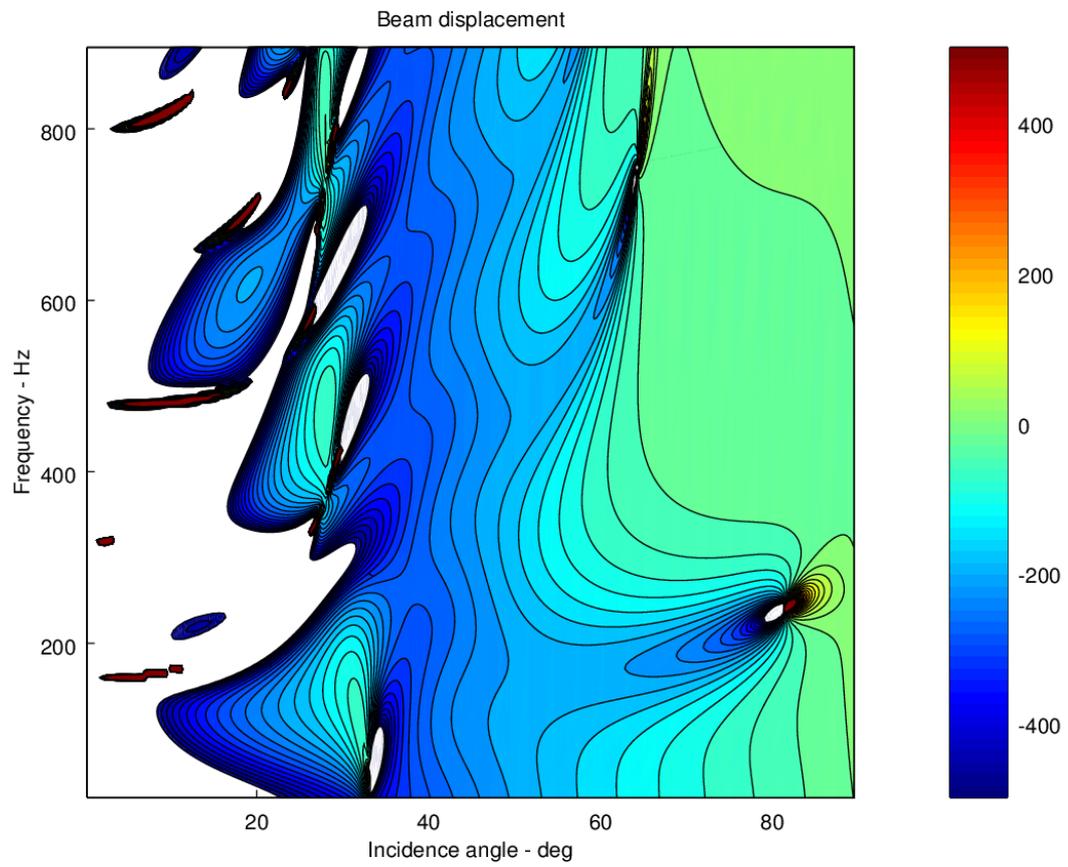

Figure 11: T4, Beam displacement (m). Both terms in Eq. 32 are included.

The reflection coefficient as mapped versus incidence angle and frequency, displays a number of dips where the magnitude is reduced. By plotting the reflection coefficient as a type of dispersion diagram in terms of phase speed versus frequency, together with the dispersion diagram for free Lamb modes it appears that the dips are closely related to these modes, indicating that the dips are caused by transfer of energy from the incident wave field into Lamb waves propagating in the ice sheet.

Shock [6] gives a formula for Lamb modes in a plate loaded in a fluid on both sides. If the fluid density is made to vanish the formula gives the free Lamb modes. Attempts with increasing fluid density does not provide "leaky" Lamb waves useful for the ice sheet, probably due to the asymmetric loading of the ice. However, a dispersion diagram for the ice sheet is obtained by Eq. 23. A contour plot shows the location of the modes along valleys in this plot. The locus of the modes seem to be deformed by the loading, and no modes can be distinguished with phase speeds larger than the longitudinal wave speed in the ice, $c_L$. On the other hand, the matrix method computation does not provide phase speeds below the sound speed in water, $c$. In the dispersion plot there also appears dips. These are evidently not directly related to the dips in the reflection map.

If absorption is excluded from the computation the reflection coefficient is of magnitude 1 almost all over; the dips shrink to very small regions.

Beam displacement is due to dependence on reflection coefficient phase angle on incidence angle ("angular term") and frequency ("frequency term"). The last term turns out to result in large negative displacements at near to normal incidence angles, which in practice is assumed to be unrealistic in a mono-frequency situation.

## 6 Acknowledgments

This work was supported by the project "Under Ice" at Nansen Environmental and Remote Sensing Center in Bergen, Norway. This report is also published as NERSC Technical Report no. 367, 2016.

## 7 Appendix

### 7.1 A Matlab code for Brekhovskikh's analytical model, Eq. 24.

```
1  % GLW.m script for reflection from ice, 8/5-2015 HH
2  testdatais % load ice parameters
3  kk=1;
4  for f=20:5:900
5  omega=2*pi*f;
6  k=omega/(cL(2)-i*aL(2));
7  K=omega/(cs(2)-i*as(2));
8  m=1;
9  for theta=0.02:.2:89
10 st2=(cL(2)-i*aL(2))/cL(1)*sind(theta);
11 st3=(cs(2)-i*as(2))/cL(1)*sind(theta);
12 ct2=sqrt(1-st2^2);
13 ct3=sqrt(1-st3^2);
14 p=k*h(2)*ct2;
```

```matlab
15  q=K*h(2)*ct3;
16  Z2=rho(2)*cL(2)/ct2;
17  ZT=rho(2)*cs(2)/ct3;
18  Z3=rho(1)*cL(1)/cosd(theta);
19  c2g=(ct3^2-st3^2)^2;
20  s2g=(2*st3*ct3)^2;
21  MZ1=Z2*c2g*cot(p)+ZT*s2g*cot(q);
22  NZ1=Z2*c2g/sin(p)+ZT*s2g/sin(q);
23  AA=MZ1^2-NZ1^2;P(m)=AA;
24  BB=MZ1*Z3;Q(m)=BB;
25  V(m)=(-BB+i*AA)/(BB+i*AA);
26  m=m+1;
27  end
28  RR(kk,:)=V(:);
29  kk=kk+1;
30  end
```

## 7.2 A Matlab code for computing Lamb modes

```matlab
1   %script Lamb6 - directly from A. Schock
2   f=20:5:2000;
3   w=2*pi*f;
4    cL=3200;%2800;
5    cS=1600;%1526;
6   cf=1500;
7   R=0;%1000/939;%1000/1119;% density ratio
8   c=(10:5:3*cS);%
9   d=5;
10  s=(cS./c).^2;
11  q=(cS/cL)^2;
12  r=(cS/cf)^2;
13  p=w*d/2./cS;%
14
15  A=tan(sqrt(q-s)'*p);%=tan(\alpha*h/2)
16  B=tan(sqrt(1-s)'*p);%=tan(\beta*h/2)
17  AA=cot(sqrt(q-s)'*p);
18  BB=cot(sqrt(1-s)'*p);
19  a1=(1-2*s).^2;
20  a2=4*s.*sqrt(1-s).*sqrt(q-s);
21  hs=a1.*AA'+a2.*BB';
22  ha=a1.*A'+a2.*B';
23  C=ones(length(f),1)*R*sqrt((q-s)./(r-s));
24  Hs=hs-i*C;
25  Ha=ha+i*C;
26
27  [H,G]=find(abs(hs)<0.015);
28  [I,J]=find(abs(ha)<0.015);
29  [na,nb]=find(abs(Hs)>1000);
30  Hs(na,nb)=-1;
31  [na,nb]=find(abs(Ha)>1000);
32  Ha(na,nb)=-1;
33
34  [t,u]=find(abs(Hs)<0.05);
35  [v,x]=find(abs(Ha)<0.05);
36  figure(1)
37  plot(f(H),c(G),'r.')
```

```
38  axis([0 2000 0 5000])
39  title('Symmetric unloaded')
40  xlabel('Frequency - Hz')
41  ylabel('Phase velocity - m/s')
42  figure(2)
43  plot(f(I),c(J),'.')
44  title('Asymmetric - unloaded - disp.')
45  axis([0 2000 0 5000])
46  xlabel('Frequency - Hz')
47  ylabel('Phase velocity - m/s')
48  figure(3)
49  plot(f(H),c(G),'r.',f(I),c(J),'.')
50  xlabel('Frequency - Hz')
51  ylabel('Phase velocity - m/s')
52  title('Symmetric: red, Asymmetric: blue, unloaded, Lamb6')
53  xlabel('Frequency - Hz')
54  ylabel('Phase velocity - m/s')
```

# References

[1] W. Thomson, "Transmission of elastic waves through stratified medium," *J. Applied Physics*, vol. 21, pp. 89–93, 1950.

[2] N. Haskell, "The dispersion of surface waves on multilayered media," *Bull. Seismol. Soc. Am*, vol. 43, pp. 17–34, 1953.

[3] L. Brekhovskikh, *Waves in Layered Media*. Academic Press, New York, USA, 1980.

[4] H. Hobæk, "Sound reflection from layered media," *NERSC Technical Report no. 340*, 2015.

[5] K. Fuchs, "Das reflexions- und transmissionsvermögen eines geschichteten mediums mit beliebiger tiefen-verteilung der elastichen moduln und der dichte für schrägen einfall ebener wellen," *Zeitschrift für Geophysik*, vol. 34, pp. 389–413, 1968.

[6] A. Schock, "Der schalldurchgang durch platten," *Acustica*, vol. 2, pp. 1–17, 1952.

[7] R. P. G. Jin, J.F. Lynch and P. Washams, "Effects of sea ice cover on acoustic ray travel times, with applications to the greenland sea tomography experiment," *J. Acoust. Soc. Amer.*, vol. 94, pp. 1044–1056, 1993.

[8] A. D. P. Alexander and N. Bose, "Modelling sound propagation under ice using the ocean acoustics library's acoustic toolbox," *Proceedings of Acoustics 2012 - Fremantle, 21-23 November 2012, Fremantle, Australia*, pp. 1–7, 2012. Australian Acoustical Society.

[9] J. D. S.D. Rajan, G.V. Frisk and C. Sellers, "Determination of compressional wave and shear wave speed profiles in sea ice by crosshole tomography - theory and experiment," *J. Acoust. Soc. Amer.*, vol. 93, pp. 721–738, 1993.